\documentclass[aps,prl,twocolumn,superscriptaddress]{revtex4}
\usepackage{graphicx}

\begin{document}


\title{First experimental determination of the charge radius of $^{74}$Rb and its application in tests of the unitarity of the CKM matrix}



\author{E.~Man\'{e}}\email[]{Ernesto@triumf.ca}\affiliation{TRIUMF, 4004 Wesbrook Mall, Vancouver, British Columbia,  V6T 2A3, Canada}
\author{A.~Voss}\affiliation{TRIUMF, 4004 Wesbrook Mall, Vancouver, British Columbia,  V6T 2A3, Canada}\affiliation{School of Physics and Astronomy, The University of Manchester, Manchester, M13 9PL, United Kingdom}
\author{J.~A.~Behr}\affiliation{TRIUMF, 4004 Wesbrook Mall, Vancouver, British Columbia,  V6T 2A3, Canada}\affiliation{University of British Columbia, Vancouver, British Columbia,  V6T 1Z1, Canada}
\author{J.~Billowes}\affiliation{School of Physics and Astronomy, The University of Manchester, Manchester, M13 9PL, United Kingdom}
\author{T.~Brunner}\affiliation{TRIUMF, 4004 Wesbrook Mall, Vancouver, British Columbia,  V6T 2A3, Canada}\affiliation{Physik Department E12, Technische Universit{\"a}t M{\"u}nchen, James-Frank-Str. 1, 85748 Garching, Germany}
\author{F.~Buchinger}\affiliation{Physics Department, McGill University, Ernest Rutherford Building, 3600 University St., Montr\'{e}al, Qu\'{e}bec, H3A 2T8, Canada}
\author{J.~E.~Crawford}\affiliation{Physics Department, McGill University, Ernest Rutherford Building, 3600 University St., Montr\'{e}al, Qu\'{e}bec, H3A 2T8, Canada}
\author{J.~Dilling}\affiliation{TRIUMF, 4004 Wesbrook Mall, Vancouver, British Columbia,  V6T 2A3, Canada}\affiliation{University of British Columbia, Vancouver, British Columbia,  V6T 1Z1, Canada}
\author{S.~Ettenauer}\affiliation{TRIUMF, 4004 Wesbrook Mall, Vancouver, British Columbia,  V6T 2A3, Canada}\affiliation{University of British Columbia, Vancouver, British Columbia,  V6T 1Z1, Canada}
\author{C.~D.~P.~Levy}\affiliation{TRIUMF, 4004 Wesbrook Mall, Vancouver, British Columbia,  V6T 2A3, Canada}
\author{O.~Shelbaya}\affiliation{Physics Department, McGill University, Ernest Rutherford Building, 3600 University St., Montr\'{e}al, Qu\'{e}bec, H3A 2T8, Canada}
\author{M.~R.~Pearson}\affiliation{TRIUMF, 4004 Wesbrook Mall, Vancouver, British Columbia,  V6T 2A3, Canada}

\date{\today}

\begin{abstract}
Collinear-laser spectroscopy with bunched-beams technique was used for the study of neutron deficient Rb isotopes, out to $^{74}$Rb ($N=Z=37$) at TRIUMF. The measured hyperfine coupling constants of $^{76,78m}$Rb were in agreement with literature values. The nuclear spin of $^{75}$Rb was confirmed to be $I=3/2$, and its hyperfine coupling constants were measured for the first time. The mean-square charge radius of $^{74}$Rb was determined for the first time. This result has improved the isospin symmetry breaking correction term used to calculate the $\mathcal{F}t$ value, with implications for tests of the unitarity of the Cabibbo-Kobayashi-Maskawa matrix.
\end{abstract}


\maketitle

The unitarity of the Cabibbo-Kobayashi-Maskawa (CKM) matrix is one of the cornerstones of the Standard Model of particles and fields. Therefore, the value of all its matrix elements should be measured as accurately and as precisely as possible for the search of new physics phenomena, such as the existence of right-handed currents, extra $Z$ bosons or another quark generation ~\cite{Hardy1}.  Indeed, over the years, the value of these matrix elements have been independently determined from different experiments. Superallowed Fermi transitions of nuclear beta decays ($0^{+} \to 0^{+}$) provide the most competitive method to determine the up-down quark matrix element $V_{ud}$ ~\cite{Yao1}. Assuming the Conserved Vector Current (CVC) hypothesis, which states that the vector coupling constant of semileptonic weak interactions $G_V$ remains unchanged in all superallowed beta transitions, this matrix element can be obtained from $V_{ud}=G_V/G_F$, where $G_F$ is the weak interaction constant for leptonic muon decay, and $G_V$ is defined in terms of the corrected $\mathcal{F}t$ value of the beta decay ~\cite{Towner4},
\begin{equation}
{\mathcal{F}t} \equiv ft(1+\delta_R^{\prime})(1+\delta_{NS}-\delta_C)=\frac{K}{2G_V^2(1 + \Delta_R^V)}.
\label{Vud}
\end{equation}
\noindent The $ft$ value is related to the measured half-lives, branching ratios and the $Q$ value of the beta decay. The terms $\delta_R^{\prime}$ and $\delta_{NS}$ are transition-dependent theoretical corrections, and $\delta_C$ is the isospin symmetry breaking term ~\cite{Towner2}. The latter is purely nuclear structure dependent, the error of which normally dominates the uncertainty on the $\mathcal{F}t$ value. The term $\Delta_R^V$ is transition-independent and $K$ is a constant. Evaluations of the isospin symmetry breaking correction involve the $13$ most precisely measured superallowed beta emitters which span a wide $Z$ range  ($ 6 \leq Z \leq 37 $) ~\cite{Towner2}. Within the existing theoretical approaches used to determine $\delta_C$ ~\cite{Towner3} the leading source of uncertainty stems from the incomplete knowledge of the radial extension of the charge distribution of the parent and daughter isotopes. Where experimental data are unavailable, the nuclear structure input is based on extrapolations of the existing models. For the lighter elements, shell model calculations with a Saxon-Woods potential provide a more consistent framework from which this information can be reliably extracted. However, with higher $Z$, the orbitals of the $fp$ shell are filled and shell model calculations become computationally difficult, due to the larger model space required. In addition, effective interactions become less reliable ~\cite{Ormand1}. There have been many theoretical efforts to fine-tune the shell model calculations in the $fp$ shell, and the importance of the $g_{9/2}$ orbital in this region has been acknowledged ~\cite{Hardy2}. Only by testing and refining these calculations against experimental data can a theoretical framework be developed for use in the calculation of $\delta_C$. In global evaluations of the $\mathcal{F}t$ value ~\cite{Towner2}, the $^{74}$Rb ($N=Z=37$) isotope is the heaviest case included. It has the biggest $\delta_C$ value among the existing cases (due to the $Z$ dependence), and the uncertainty around the $\delta_C$ dominates the total uncertainty of the $\mathcal{F}t$ value for $^{74}$Rb at approximately $30\%$ when all the fractional uncertainties are combined ~\cite{Hardy1}.

The isotope shift of the hyperfine atomic spectrum is related to changes in the mean square charge radius ~\cite{Cheal1}. The hyperfine structure allows one to extract the magnetic dipole moment of the nucleus, $\mu$, if the nucleus has non-zero nuclear spin and if there exists a calibration isotope with a known moment. If the spin is $I>1/2$ the spectroscopic electric quadrupole moment, $Q_s$ can in principle be measured. Furthermore, the hyperfine structure allows for a true measurement of the nuclear spin. High resolution laser spectroscopy is a powerful tool for probing the hyperfine atomic spectra, and the collinear method with fast beams has the sensitivity and resolution required for radioisotopes ~\cite{Cheal1}. On the other hand, low production ($\sim 10^4$ ions/s) and background have become limiting factors for the sensitivity of this type of experiments. The use of cooler-buncher traps in conjunction with the collinear method is an elegant solution to this problem, as demonstrated at the IGISOL ~\cite{Nieminen1}, and ISOLDE ~\cite{Mane2} facilities. The ISAC facility houses TRIUMF's Ion Trap for Atomic and Nuclear science (TITAN)~\cite{Dilling1}. This setup comprises a series of ion trapping devices used in particular for high precision mass measurements and decay spectroscopy studies. The first of these devices is a segmented linear Paul trap which is placed vertically in the ISAC beam line ~\cite{Smith1}. This device can deliver cooled and bunched beams in forward, and uniquely, in reverse mode. The successful integration of the reverse-extraction of radioactive bunched ions for the collinear laser spectroscopy program has been demonstrated in the facility and is described in detail in ~\cite{Mane1}. In this Letter, we present the first application of this technique for the determination of the charge radius of $^{74}$Rb. This result constitutes the first experimental benchmark of the theoretically determined $\delta_C$. 

Radioactive rubidium atoms were produced at the ISAC facility from spallation of a stack of natural Nb foils (density  $\rho \sim 8.5$~g$/$cm$^3$) irradiated by $98\ \mu$A of protons at $500$ MeV from TRIUMF's cyclotron. The Rb atoms were surface-ionized, extracted at $28$ keV, mass-separated and sent to the main experimental hall. The yield of $^{74}$Rb measured at a tape station was $1.7\times\ 10^4$ ions$/$s. The ions were injected into TITAN's cooler-buncher and subsequently reverse-extracted to the laser spectroscopy setup, as described in reference ~\cite{Mane1}. The ion beam was neutralized in a charge-exchange cell filled with Na vapor heated to $445\ ^{\circ}$C (neutralization efficiency of $50\%$). A frequency stabilized diode laser tuned to the $5s\ ^{2}S_{1/2} \rightarrow 5p\ ^{2}P_{3/2}$ transition, the ``$D_2$ line'' ($780$ nm) was used, with $1.5$ mW of light coupled to the beam line via an optical fiber, focused to approximately $1.5$ mm diameter, and overlapped the atomic beam in a collinear geometry. For this purpose, two retractable apertures ($1$ mm and $3$ mm diameter) separated by approximately $12.5$ cm were introduced in the detection region. The frequency locking of the diode laser was based on a polarization stabilized He-Ne laser, using the principle described in reference ~\cite{Mane1}. The ion beam was decelerated by a voltage applied to the charge-exchange cell and Doppler-shifted into resonance. The laser induced fluorescence at $90^{\circ}$ was recorded with a red-sensitive photomultiplier tube. An interference filter was used to help reduce the detected laser scatter. The cooler-buncher trap operated with a repetition rate of $10$ Hz. Within this $100$ ms cycle, $80$ ms was used to accumulate the ions into the trap and $1$ ms was used for cooling. The trap was left open for $100$ $\mu$s for releasing the ions, and the extracted bunches had a temporal length of $4$ $\mu$s at its full width at half maximum (FWHM). A $12$ $\mu$s hardware gate was set on the detected photon signal in coincidence with the arrival of the ion bunch. The background suppression achieved was four orders of magnitude.
 
\begin{figure*}[ht]
\includegraphics[height=.45\textheight]{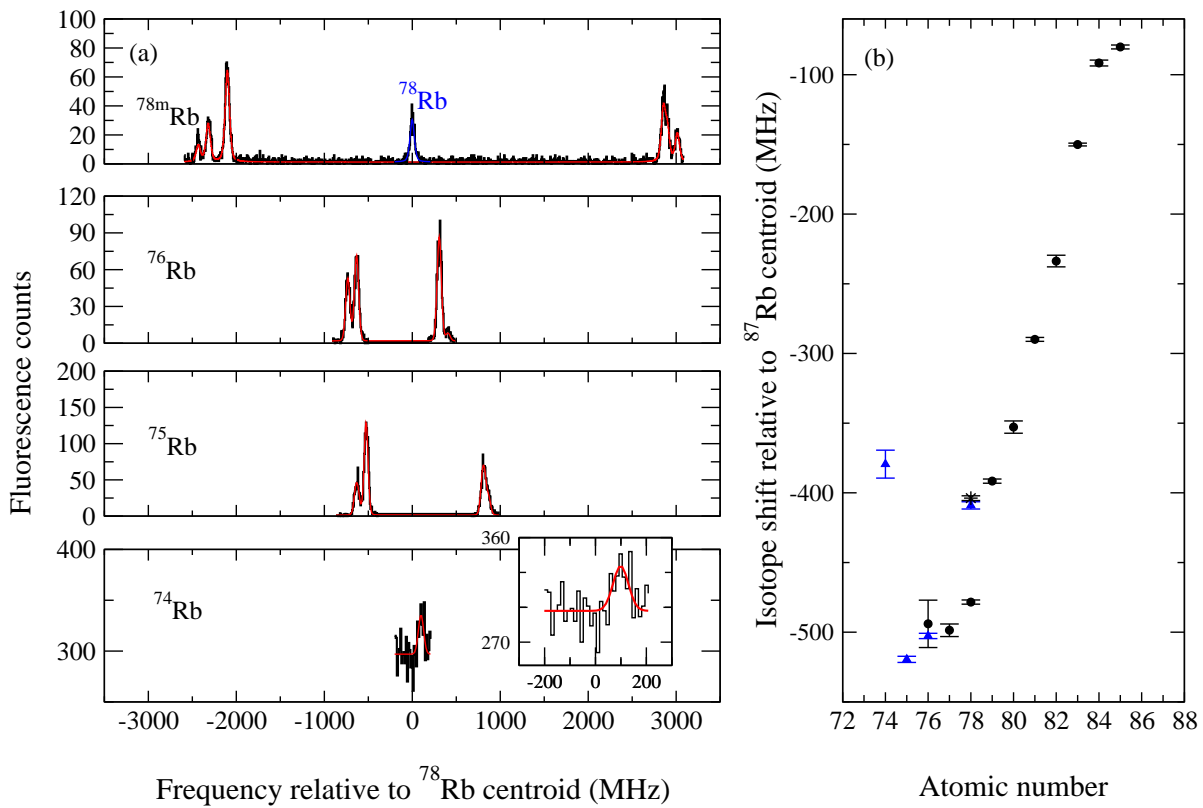}
\caption{\label{fig:allspectra}(Color on-line)(a)  Optical spectra and fits for the neutron deficient Rb isotopes, measured on the $D_2$ line, relative to $^{78}$Rb. The insert is an expanded plot of the $^{74}$Rb dataset. (b) Isotope shifts of rubidium isotopes relative to $A=87$. The black circles indicate the dataset presented in ~\cite{Thibault1}, with the black asterisk indicating $\delta\nu^{78m-87}$Rb. The blue triangles are the isotope shifts measured in this work.}
\end{figure*}

The hyperfine structure of both the ground state ($I=0$, $T_{1/2}=17.66$ m) and the isomeric state ($E=103$ keV, $I=4$, $T_{1/2}=5.74$ m) of $^{78}$Rb are shown at the top of Figure~\ref{fig:allspectra}.(a).
\noindent The single peak at the center of the figure corresponds to the ground state spin of $^{78}$Rb ($I=0$, and hence exhibits no hyperfine splitting). A $\chi^2$ minimization routine was used for fitting the spectra using a Voigt profile. The same figure shows all the six hyperfine transitions of the isomeric state. For the hyperfine structure of the isomer, the background and lineshape were fixed to the values obtained from the fit of the ground state transition. The plot also shows the hyperfine structure of $^{76}$Rb together with the fit. As for $^{75}$Rb, its measured hyperfine structure permitted an unambiguous ground-state spin assignment. Previously, the ground-state spin for this isotope had only been tentatively assigned as either $I=3/2$ or $I=5/2$ based on beta and gamma decay studies ~\cite{Kern1,Gross1}. A comparison of the $\chi^2$ between the two assumptions confirms the $I=3/2$ assignment. The bottom of Figure~\ref{fig:allspectra}.(a) shows the fluorescence spectrum of $^{74}$Rb in a $400$ MHz scanning region. We have excluded the resonance being in the region following the trend of the heavier isotopes. Since this isotope has ground-state spin $I=0$, only one peak is observed. The signal-to-noise ratio of this peak is $2.2$. The fitted FWHM is $60(17)$ MHz, which agrees with the systematics of the measured widths of all the measured hyperfine transitions of $^{75,76,78m,78}$Rb. The results in Table~\ref{tab:summary} summarize the hyperfine constants and the isotope shifts of the Rb isotopes measured in this work. The hyperfine coupling constants obtained for $^{76,78m}$Rb, and the isotope shifts relative to $^{78}$Rb confirm the previously measured results reported in ~\cite{Thibault1}.  
\begin{table}[h]
\caption{\label{tab:summary}Summary of the isotope shifts and hyperfine coupling constants (in MHz) of the Rb isotopes measured in the $5s\ ^{2}S_{1/2} \rightarrow 5p\ ^{2}P_{3/2}$ transition. For $^{75}$Rb, the results of the hyperfine coupling constants obtained from the fits allowing for the two spin possibilities are shown.}
\begin{ruledtabular}
\begin{tabular}{ccccccc}
$A$ & $I$    & $\delta\nu^{A-78}$ & $A(S_{1/2})$  & $A(P_{3/2})$ & $B(P_{3/2})$ & $\chi^2_r$  \\
\hline
78  &  $0$     & $0$ 	        &    $-$        &     $-$      &    $-$         & $0.83$   \\
78m &  $4$   & $+69.4(21)$        & $+1185.1(05)$ & $+29.3(01)$  &   $+83.1(22)$ & $0.76$     \\
76  &  $1$   & $-24.3(12)$        & $-693(08)$    & $-17.15(02)$ &   $+32(07)$   & $1.08$    \\
75  &  $3/2$ & $-41.1(17)$        & $+719.6(10)$  & $+17.8(01)$  &   $+63(27)$   &  $1.18$    \\
    &  $5/2$ & $+12.1(12)$        & $+478.1(10)$  & $+11.9(02)$  &   $+85.5(22)$ & $1.87$    \\
74  &  $0$     & $+99(10)$ 	        &    $-$        &     $-$      &    $-$ &  $0.93$          \\
\end{tabular}
\end{ruledtabular}
\end{table}
The isotope shifts $\delta\nu^{AA^{\prime}}$ were determined from the relative differences of the centroid frequencies between the isotopes ($A$) and the isotope of reference ($A^{\prime}$), $^{78}$Rb. Before and during the data collection of $^{74,75,76}$Rb, calibration measurements of $^{78}$Rb were performed and the system has shown to be stable to $1$ MHz/day. Figure ~\ref{fig:allspectra}.(b) shows the isotope shifts of rubidium isotopes relative to $A=87$. It is interesting to point out the significant deviation of $^{74}$Rb from the trend.

The theory of optical isotope shifts separates the observed change in the transition frequency between two different isotopes with atomic masses $A$ and $A^{\prime}$ into the mass shift and field shift terms,
\begin{eqnarray}
\delta\nu^{AA^{\prime}} &=& \delta\nu_{MS}^{AA^{\prime}} + \delta\nu_{FS}^{AA^{\prime}}{}\nonumber\\
                      {}&=& K \left( \frac{A^{\prime}-A}{AA^{\prime}} \right) + F\delta \langle r_{ch}^2 \rangle^{AA^{\prime}}.
\label{isotopeshift}
\end{eqnarray}

\noindent The mass shift constant $K$ is separable into the normal and the specific mass shift. The first is a reduced mass effect and is exactly calculable, whereas the second is an effect of the change in the correlations between pairs of electrons for the different isotopes. The quantity $F$ is the field shift constant and depends on the atomic structure. In the case of the $D_2$ transition of rubidium, recent calculations performed by Dzuba-Johnson-Safronova produce $F=-567.45$ MHz/fm$^2$ with $1\%$ uncertainty ~\cite{Dzuba1}. Since the isotope shifts establishes a relationship with \textit{changes} in the mean square charge radius, the charge radius of a usually stable reference needs to be used in order to extract an absolute value for the radius of the isotope of interest. The relative rms charge radii of $^{85,87}$Rb have been measured by muonic X-rays to be $^{85}\text{Rb}:\langle r_{ch}^2 \rangle^{1/2} = 4.2035(3)$ fm and $^{87}\text{Rb}$ (reference isotope) $:\ \langle r_{ch}^2 \rangle^{1/2} = 4.1985$ fm ~\cite{Angeli1}, such that $\delta \langle r_{ch}^2 \rangle^{85,87}=0.042(2)$ fm$^2$. With the information that $\delta\nu^{85-87}=-77.992(20)$ MHz at the $D_2$ transition ~\cite{Banerjee1}, the mass shift constant on Equation \ref{isotopeshift} is $K=-200(12)$ GHz$/$amu. The work of Thibault \textit{et al.} gives $\delta\nu^{78-87}=-478.4(15)$ MHz at the $D_2$ transition ~\cite{Thibault1}. Therefore, it is straightforward to tie this result together with our finding presented in Table \ref{tab:summary} in order to obtain the isotope shift between $^{74}$Rb and $^{87}$Rb, viz. $\delta\nu^{74-87}=-379(10)$ MHz. Inserting this value into Equation \ref{isotopeshift} gives $\delta \langle r_{ch}^2 \rangle^{74,87}=-0.04(5)$ fm$^2$, and therefore the rms charge radius of $^{74}$Rb is determined to be  $\langle r_{ch}^2 \rangle^{1/2} = 4.19(1)$ fm. The need for microscopic calculations to explain this result is motivated by the fact that admixtures of two $I^{\pi}=0^{+}$ configurations with prolate/oblate deformations have been observed in the daughter nucleus $^{74}$Kr. The strength of the $E0$ transition between these states was used as input to quantitatively understand how two-state admixtures are manifest in the charge radius~\cite{Wood1}. The observation of the isomeric decay of $^{74}$Kr by Chandler \textit{et al.} has led the authors to suggest that the ground state wavefunction is composed of admixtures of states of opposite quadrupole deformation $\beta_2 = -0.35$ and $\beta_2 = +0.38$. The authors suggested that the mixing amplitude should lie between $30-50\%$ ~\cite{Chandler1}. On the other hand, there has been so far no experimental evidence in support of a competing  $0^{+}$ state in $^{74}$Rb ~\cite{Rudolph1,Fischer1}. Shape coexistence and proton-neutron ($pn$) pairing correlations are expected to play a role in the $N \approx Z$ region ~\cite{Warner1}, and the isospin dependence of the pairing interaction is likely to affect the excitation energy of the first excited $0^{+}$ state. Since the charge radius is sensitive to static and dynamic deformation effects, the result for $^{74}$Rb reported in this work can be used to assess the composition of the ground-state wavefunction. 

The main impact of the measurement presented in this work lies in the reevaluation of the isospin symmetry correction term. In the analysis performed by Towner and Hardy ~\cite{Towner4} the authors split $\delta_C$ into two contributions, $\delta_C = \delta_{C1} + \delta_{C2}$. The first term $\delta_{C1}$ reflects differences in isospin mixing between the parent and daughter isotopes, whereas the second term $\delta_{C2}$ accounts for the radial overlap correction between parent and daughter isotopes. Although $\delta_{C1}$ is sensitive to the charge dependence of the effective interaction, its contribution is approximately $10\%$ of $\delta_{C2}$. The published value of $\delta_{C2}$ for $^{74}$Rb ~\cite{Towner2} uses an extrapolation for the rms charge radius ($\langle r_{ch}^2 \rangle^{1/2} = 4.18(10)$ fm). This value was then used to fix the radius parameter in the Saxon-Woods central potential in order to calculate the radial-mismatch factor. For the computation of $\delta_{C2}$ the error budget comprises three pieces - ($1$) the component related to the mean square charge radius, $0.15$; ($2$) the spread from Methods II, III and IV, $0.20$ (see ref. \cite{Towner2}) and ($3$) the spread among the effective interactions considered, $0.12$. Combining in quadrature and rounding the result gives the error of $0.30$, viz. $\delta_{C2}=1.50(30)\%$.  With the measured mean square charge radius of $^{74}$Rb presented in this work, the first error reduces by an order of magnitude, while the other two error components remain unchanged. This results in a reduction of the error on $\delta_{C2}$ by $\sim 20\%$, and hence reduces the uncertainty on the corresponding $\delta_C$ value. When one considers the error budget for the $\mathcal{F}t$ value, this result puts the uncertainty on $\delta_C$ to levels comparable to the uncertainty of the $Q$-value, which is the next leading term ~\cite{Herfurth1,Kellerbauer1,Kellerbauer2}. Therefore, our result stimulates further experimental work to reduce the uncertainty of the latter. Indeed, this work has motivated recent precision Penning trap mass measurements on highly charged $^{74}$Rb performed at TRIUMF. 

In summary, this work reports on the first measurement of extremely neutron deficient rubidium isotopes by optical methods. A novel experimental approach was developed and used to achieve unprecedented levels of sensitivity in the measurements of short-lived and low intensity radioisotopes produced at ISAC. The information obtained on the charge radius of $^{74}$Rb was used to reduce the uncertainty in the evaluation of the isospin symmetry breaking correction term. This is the first time a charge radius measurement is being used for validating this correction, in what constitutes one step towards improving the precision and accuracy of $V_{ud}$, for tests of the unitarity of the CKM matrix. This experimental program will now be extended to measure the charge radius of $^{62}$Ga, the second heaviest superallowed beta emitter. 

This work has been supported by NSERC and NRC through TRIUMF. The authors would like to thank Prof. I.~Towner for providing the recalculated value of $\delta_{C2}$ and G. Ball for the useful comments. We would also like to thank D. Bishop, S. Daviel and M. Good for their support and assistance. J.~A.~Behr would like to thank B. Lee for help in an earlier attempt at this experiment. J.~B. and A.~V. acknowledges support from the STFC of the United Kingdom.

\bibliography{myref}

\end{document}